\documentclass{article}

\usepackage{graphicx}
\usepackage{biblatex}
   \usepackage[nonatbib, final]{neurips_2023}

\usepackage[utf8]{inputenc} 
\usepackage[T1]{fontenc}    
\usepackage{hyperref}       
\usepackage{url}            
\usepackage{booktabs}       
\usepackage{amsfonts}       
\usepackage{nicefrac}       
\usepackage{microtype}      
\usepackage{xcolor}         

\title{Democratise with Care: The need for fairness specific features in user-interface based open source AutoML tools}

\usepackage{authblk}

\author{%
\quad Sundaraparipurnan Narayanan\\
\texttt{Doctoral Candidate, University of Bordeaux}\\
\texttt{sundar.narayanan@aitechethics.com}
}

\begin{document}

\maketitle

\begin{abstract}
 AI is increasingly playing a pivotal role in businesses and organizations, impacting the outcomes and interests of human users. Automated Machine Learning (AutoML) streamlines the machine learning model development process by automating repetitive tasks and making data-driven decisions, enabling even non-experts to construct high-quality models efficiently. This democratization allows more users (including non-experts) to access and utilize state-of-the-art machine-learning expertise. However, AutoML tools may also propagate bias in the way these tools handle the data, model choices, and optimization approaches adopted. We conducted an experimental study of User-interface-based open source AutoML tools (DataRobot, H2O Studio, Dataiku, and Rapidminer Studio) to examine if they had features to assist users in developing fairness-aware machine learning models. The experiments covered the following considerations for the evaluation of features: understanding use case context, data representation, feature relevance and sensitivity, data bias and preprocessing techniques, data handling capabilities, training-testing split, hyperparameter handling, and constraints, fairness-oriented model development, explainability and ability to download and edit models by the user. The results revealed inadequacies in features that could support in fairness-aware model development. Further, the results also highlight the need to establish certain essential features for promoting fairness in AutoML tools.
 
\end{abstract}

\section{Introduction}
Artificial intelligence (AI), a disruptive technology since its inception, impacts various human rights issues, from discrimination to supply chains \cite{Kriebitz2020} . AI isn't just for engineers and technical staff; it's a business necessity. As technology evolves over the coming years, more significant risks will be associated with adopting advanced technology\cite{Bartneck2021} . AI is increasingly playing a pivotal role in businesses and organizations, impacting the outcomes and interests of human users \cite{david2021} . Many machine learning methods' performance is susceptible to many design decisions. To achieve consistent success, machine learning experts must select the appropriate features, workflows, and algorithms; these complex tasks may be too difficult for non-experts to tackle alone, spurring demand for methods that can be employed without special knowledge or skillsets. Such methods are collectively called Automated Machine Learning \cite{Cham2019}.

Automated Machine Learning, also called automated ML or AutoML automates the tedious, repetitive tasks involved in machine learning model development. This enables data scientists, analysts, developers, and programmers to construct large-scale models that are efficient and productive while still maintaining model quality. Automated machine learning (AutoML) automates these decisions through data-driven, objective methods: the user provides input data and the AutoML system automatically determines the most suitable approach for their particular use case. AutoML is perceived as a democratization of machine learning; with AutoML, customized state-of-the-art machine learning expertise is within everyone's reach. 

Historic adverse incidents have established that AI technology also carries risks that could negatively affect individuals, groups and organizations, communities, society, and the environment. With the increasing importance of algorithmic decision-making in distributing limited resources and the growing dependence of humans on AI's outcomes, concerns regarding fairness are on the rise. Evidence shows that algorithmic systems can perpetuate racism, classism, sexism, and other forms of discrimination that cause tangible harm. They were implicated in cases where people of color were denied benefits such as kidney transplants and mortgages due to biases in facial recognition technology \cite{Saha2022}. These are just some documented harm caused by algorithmic systems \cite{markup}.

Essentially, Fairness in ML ensures that ML models do not discriminate against certain groups of people. For example, a fairness-aware ML model should not predict that a person is more likely to commit a crime based on race or gender. AutoML systems are not an exception in this context. AutoML systems are becoming more widely utilized. Gartner predicts that sixty-five percent of all application development activities will use low-code or no-code tools by 2024. AutoML falls in the low-code or no-code applications category, enabling non-expert users to build machine learning models. Non-expert users of AutoML tools may be particularly vulnerable to fairness concerns. This is because they may not have the expertise to identify or address potential fairness issues in the data or the model. Hence, assessing the risks of unfair AutoML systems and whether there are sufficient fairness mitigation techniques to increase overall fairness is essential \cite{Wu}. 
This paper deals with the experimental study of four AutoML tools to examine if they sufficiently support fairness expectations for non-expert users. This research is focused on gathering information on features within the AutoML tools that help fairness. 

\section{Experimental Study}
\subsection{Identification of AutoML tools for the experiment}

To identify AutoML tools for fairness features evaluation, we reviewed 16 tools, including 10 user interface (UI)-based tools (RapidMiner Studio, H2O Studio, Google Vertex AI, Amazon Web Services AutoML, IBM WatsonX, Azure AutoML, Data Robot, Dataiku, KNIME, and Alteryx) and 6 code libraries (PyCaret, Auto-SKlearn, AutoGluon, MLBOX, TPOT, and Auto-Keras). We focused on UI-based tools because they are more accessible to users of all skill levels. Of the UI-based tools, we excluded cloud-based platform tools that require platform understanding for use (Google Vertex AI, Amazon Web Services AutoML, IBM WatsonX, and Azure AutoML). For the above identification, we considered the following criteria: (1) Ease of use: AutoML tools should be easy to learn and use, even for users with limited machine learning experience; (2) Data loading and preparation: AutoML tools should provide a simple and efficient way to load and prepare data for machine learning; (3) Platform accessibility: AutoML tools should be accessible to users on a variety of platforms, including cloud-based and desktop environments; and (4) Non-expert friendly user interface: AutoML tools should have a user interface that is easy to understand and use, even for users with no prior knowledge of machine learning. These criteria were chosen to ensure that we selected AutoML tools that are accessible and useful to a wide range of users. This left us with six UI-based tools for evaluation: RapidMiner Studio, H2O Studio, Data Robot, Dataiku, KNIME, and Alteryx. For the research, we examine RapidMiner Studio, H2O Studio, Data Robot, and Dataiku, in the first phase.

\subsection{Fairness features consideration for experimental study}

This experimental study focused on an audit approach to examine the fairness features available in the UI based tools for evaluation.  The fairness features considered for the experimejtal study could includes Dataset diversity and representation \cite{TrewinShari2019}, checks for biases and unbalanced data \cite{Zhang2019}, applies consistent preprocessing \cite{Mehrabi2021}, ensures fair training-testing split \cite{Mehrabi2021}, removes sensitive features \cite{Fu2020}; \cite{Ntoutsi2020}, avoids hyperparameter bias, prevents overfitting \cite{Zhang2022}, tests on diverse data, avoids underfitting \cite{Cunningham2020}, evaluates using fairness-sensitive metrics \cite{Castelnovo2021} \cite{Castelnovo2022} \cite{Mukherjee2020} \cite{Saleiro2018} and compares to a benchmark \cite{Tai2022} , alerts about high-risk sensitive features \cite{Li2022}  \cite{Veale2018}, ensures model robustness to data variations , and handles missing or incomplete, noisy or corrupted, rare or unusual data points \cite{Kallus2021} \cite{Chen2019} , or imbalanced datasets in a fair and unbiased manner \cite{Chen2022}. In addition, we considered seeking use case context, understanding data sensitivity, handling monotonic constraints, developing fairness-aware models, explaining the prediction results, handling model documentation, allowing users to download the model/ code, and highlighting the model's limitations as additional considerations for evaluation. 

\subsection{Approach towards experiments}
Our research aimed to assess the fairness support capabilities of four prominent AutoML tools: RapidMiner Studio, H2O Studio, DataRobot, and Dataiku. The study was initiated by downloading and installing the necessary software. Subsequently, a sample dataset was introduced. The dataset pertains to the direct marketing initiatives (phone calls) conducted by a banking institution in Portugal. The dataset comprised information on credit default occurrences among customers, including demographic variables such as age and marital status \cite{Moro2014}. The main aim of our study was to utilize the provided dataset to train machine learning models to predict the likelihood of a credit default. During our review, we directed our attention toward criteria to evaluate the fairness capabilities of each tool. In this study, we investigated the presence of three fundamental categories of features: (a) features that could promote fairness, (b) features that could facilitate the identification of fairness issues by the user, and (c) features that had the options for users to make choices regarding feature selection, model building, and other related processes.

\section{Results from the Study}
In exploring fairness features in prominent AutoML tools, we evaluated Rapidminer Studio, H2O Studio, Data Robot, and Dataiku across various criteria for fair machine learning development.
\begin{enumerate}
  \item \textbf{Use Case Context and Understanding}: None of the tools provided features specifically seeking the context of use cases. Similarly, no tool provided users with options for specifying use case context or understanding data sensitivity.
  \item \textbf{Data Representation}: All tools lacked explicit features for ensuring dataset diversity and representation. Additionally, they did not support user-based discovery for dataset diversity or provide options. The tools had the option to assess the statistical distribution of the features.
  \item \textbf{Feature Relevance and Sensitivity}:  Each tool successfully highlights the relevance of features, supporting user-based discovery of such elements. While Rapidminer Studio could highlight and handle sensitive features, the others did not have such a feature. 
  \item \textbf{Data Bias and Preprocessing}: Bias detection was notably weak across the tools. However, Rapidminer Studio and Dataiku allowed users to discover unbalanced data. All tools had a mechanism for consistently applying preprocessing approaches, Rapidminer Studio, H2O Studio, and Dataiku allowed users to find these preprocessing methods, while Rapidminer provided users with an option to make changes to the pre-processing parameters. 
  \item \textbf{Data Handling Capabilities}: Each tool could handle missing or atypical data points. H2O Studio, Data Robot, and Dataiku also supported user discovery for these data handling methods. However, none of the tools allowed the user to suggest/ change the approach to handling missing or atypical data on the application.  
  \item \textbf{Training and Testing Split}: All tools ensured a consistent approach to the training-testing split. However, the tools did not allow the user to determine the extent of the training and testing split, except Data Robot, which offered limited customizability based on percentages.
  \item \textbf{Hyperparameters and Constraints}: All the tools handled hyperparameters, but only Dataiku enabled user discovery for this process. Further, only H2O provided an option for the user to make changes to the hyperparameter considerations. Also, only Dataiku provided a feature for handling constraints, allowing users to discover and modify them.
  \item \textbf{Fairness-Oriented Model Development}: None of the tools had features to develop fairness-aware models. While Dataiku had a part through which the user can evaluate the model's fairness developed with specific metrics (demographic parity, equalized odds, equality of opportunity, and predictive rate parity), the other tools did not have any such feature. 
  \item \textbf{Explainability and Model Limitations}: Rapidminer Studio, Data Robot, and Dataiku could explain decision-making, with Data Robot offering explanations through an app interface supporting users to discover the model performance through explanations. RapidMiner had a feature that allowed users to choose whether to train the model to provide explanations for predictions or otherwise at the time of model development. However, none of the applications provided any limitations associated with the models for the user to be aware when deploying these models in a certain use case context.   
  \item \textbf{Documentation and Download of Model}: All tools provided documentation of the model training; however, this documentation does not include any references to fairness-aware strategies applied in the model. In addition, H2O, Dataiku, and Rapidminer allow the user to download the model for changes or deployment elsewhere. 

\end{enumerate}

\section{Discussion and Conclusion}
The evaluation of fairness attributes in prominent AutoML tools, such as Rapidminer Studio, H2O Studio, Data Robot, and Dataiku, has unveiled a disconcerting deficiency in providing full assistance for fairness-aware machine learning. Although all the tools described in this study possess qualities such as highlighting the importance of features and capacities for processing data, none of them specifically address the development of fairness. The observed deficiencies include the lack of features that address the contextual use case, promote dataset diversity, or facilitate the creation of models with a focus on fairness. The solutions focused on data processing functionalities such as managing missing data or hyperparameters but did not provide users with substantial capabilities to customize fairness strategies.

As noted, the absence of fairness features or context-aware considerations has the potential to result in the development of biased or unfair models unintentionally. Considering that these tools are frequently utilized by individuals lacking expertise in the field, an increased risk is involved. These users may lack awareness regarding the potential biases included in their models, hence perpetuating biased decision-making. The absence of fairness considerations can result in societal ramifications in a socio-technological environment where machine learning models substantially influence decision-making processes. The results underscore the need for AutoML platforms to integrate comprehensive fairness-aware features, ensuring more ethical and equitable machine learning deployments.

In conclusion, while the examined AutoML tools demonstrate robust features in some fairness-related regions, there are significant gaps, especially concerning developing fairness-aware models and evaluations. While the current research focused on fairness-aware considerations, AutoML tools, especially those designed for non-experts, shall prioritize features that promote responsible and trustworthy models to ensure their widespread adoption doesn't perpetuate harm. Some examples of these features include bias detection and mitigation techniques, built-in privacy protections, and explainability mechanisms to help users understand the model's decision-making process. Additionally, AutoML tools should incorporate ethical considerations in their optimization process, ensuring that the models produced are fair, transparent, and aligned with societal values. By combining these functionalities, AutoML can democratize machine learning but also help mitigate potential risks and foster trust in the technology.




\printbibliography

@article{Zhang2019,
  title = {Fairness Assessment for Artificial Intelligence in Financial Industry},
  author = {Yukun Zhang and Longsheng Zhou},
  year = {2019},
  url = {https://arxiv.org/abs/1912.07211v1},
}

@ARTICLE{Zhang2022,
  doi = {10.1148/RYAI.220010/ASSET/IMAGES/LARGE/RYAI.220010.FIG4.JPEG},
  journal = {Radiology: Artificial Intelligence},
  publisher = {Radiological Society of North America Inc.},
  title = {Mitigating Bias in Radiology Machine Learning: 2. Model Development},
  volume = {4},
  url = {https://pubs.rsna.org/doi/10.1148/ryai.220010},
  year = {2022},
}

@article{Wu,
   author = {Qingyun Wu and Chi Wang},
   title = {FAIRAUTOML: EMBRACING UNFAIRNESS MITIGATION IN AUTOML},
}

@article{Veale2018,
  author = {Michael Veale and Max Van Kleek and Reuben Binns},
  doi = {10.1145/3173574.3174014},
  journal = {Conference on Human Factors in Computing Systems - Proceedings},
  title = {Fairness and accountability design needs for algorithmic support in high-stakes public sector decision-making},
  url = {https://dl.acm.org/doi/10.1145/3173574.3174014},
  year = {2018},
}

@article{Kriebitz2020,
  author = {Alexander Kriebitz and Christoph Lütge},
  doi = {10.1017/BHJ.2019.28},
  journal = {Business and Human Rights Journal},
  title = {Artificial Intelligence and Human Rights: A Business Ethical Assessment},
  url = {https://www.cambridge.org/core/journals/business-and-human-rights-journal/article/abs/artificial-intelligence-and-human-rights-a-business-ethical-assessment/33D07AB42FC76A4BA49B03F600186E1B},
  year = {2020},
}

@article{Bartneck2021,
  author = {Christoph Bartneck and Christoph Lütge and Alan Wagner and Sean Welsh},
  doi = {10.1007/978-3-030-51110-4_6},
  journal = {SpringerBriefs in Ethics},
  title = {Risks in the Business of AI},
  url = {https://link.springer.com/chapter/10.1007/978-3-030-51110-4_6},
  year = {2021},
}

@article{david2021,
  author = {David De Cremer and Garry Kasparov},
  doi = {10.1007/S43681-021-00075-Y},
  journal = {AI and Ethics 2021 2:1},
  title = {The ethical AI—paradox: why better technology needs more and not less human responsibility},
  url = {https://link.springer.com/article/10.1007/s43681-021-00075-y},
  year = {2021},
}

@article{Cham2019,
  author = {Frank Hutter and Lars Kotthoff and Joaquin Vanschoren},
  doi = {10.1007/978-3-030-05318-5},
  publisher = {Springer International Publishing},
  title = {Automated Machine Learning},
  url = {http://link.springer.com/10.1007/978-3-030-05318-5},
  year = {2019},
}

@article{Saha2022,
  author = {Sasha Costanza-Chock and Inioluwa Deborah Raji and Joy Buolamwini},
  doi = {10.1145/3531146.3533213},
  journal = {ACM International Conference Proceeding Series},
  title = {Who Audits the Auditors? Recommendations from a field scan of the algorithmic auditing ecosystem},
  url = {https://dl.acm.org/doi/10.1145/3531146.3533213},
  year = {2022},
}

@article{markup,
  author = {Emmanuel Martinez and Lauren Kirchner},
  journal = {ACM International Conference Proceeding Series},
  title = {The Secret Bias Hidden in Mortgage-Approval Algorithms – The Markup},
  url = {https://themarkup.org/denied/2021/08/25/the-secret-bias-hidden-in-mortgage-approval-algorithms},
  year = {2021},
}

@article{Moro2014,
  author = {Sérgio Moro and Paulo Cortez and Paulo Rita},
  doi = {10.1016/J.DSS.2014.03.001},
  journal = {Decision Support Systems},
  title = {A data-driven approach to predict the success of bank telemarketing},
  year = {2014},
}

@article{Chen2022,
  author = {Zhenpeng Chen and Jie M. Zhang and Federica Sarro and Mark Harman},
  doi = {10.1145/3583561},
  journal = {ACM Transactions on Software Engineering and Methodology},
  title = {A Comprehensive Empirical Study of Bias Mitigation Methods for Machine Learning Classifiers},
  url = {https://arxiv.org/abs/2207.03277v3},
  year = {2022},
}

@article{Chen2019,
  author = {Jiahao Chen and Nathan Kallus and Xiaojie Mao and Geofry Svacha and Madeleine Udell},
  doi = {10.1145/3287560.3287594},
  journal = {FAT* 2019 - Proceedings of the 2019 Conference on Fairness, Accountability, and Transparency},
  title = {Fairness under unawareness: Assessing disparity when protected class is unobserved},
  url = {https://dl.acm.org/doi/10.1145/3287560.3287594},
  year = {2019},
}

@article{Kallus2021,
  author = {Nathan Kallus and Xiaojie Mao and Angela Zhou},
  doi = {10.1287/MNSC.2020.3850},
  title = {Assessing Algorithmic Fairness with Unobserved Protected Class Using Data Combination},
  url = {https://pubsonline.informs.org/doi/abs/10.1287/mnsc.2020.3850},
   year = {2021},
}

@article{Li2022,
  author = {Mingchen Li and Xuechen Zhang and Christos Thrampoulidis and Jiasi Chen and Samet Oymak},
  journal = {Advances in Neural Information Processing Systems},
  title = {AutoBalance: Optimized Loss Functions for Imbalanced Data},
  url = {https://arxiv.org/abs/2201.01212v1},
  year = {2022},
}

@article{Tai2022,
   author = {Tai Le Quy and Arjun Roy and Vasileios Iosifidis and Wenbin Zhang and Eirini Ntoutsi},
   doi = {10.1002/WIDM.1452},
   journal = {Wiley Interdisciplinary Reviews: Data Mining and Knowledge Discovery},
   title = {A survey on datasets for fairness-aware machine learning},
   url = {https://onlinelibrary.wiley.com/doi/full/10.1002/widm.1452}, 
   year = {2022},
}

@article{Saleiro2018,
   author = {Pedro Saleiro and Benedict Kuester and Loren Hinkson and Jesse London and Abby Stevens and Ari Anisfeld and Kit T. Rodolfa and Rayid Ghani},
   title = {Aequitas: A Bias and Fairness Audit Toolkit},
   url = {https://arxiv.org/abs/1811.05577v2},
   year = {2018},
}

@article{Mukherjee2020,
   author = {Debarghya Mukherjee and Mikhail Yurochkin and Moulinath Banerjee and Yuekai Sun},
   publisher = {PMLR},
   title = {Two Simple Ways to Learn Individual Fairness Metrics from Data},
   url = {https://proceedings.mlr.press/v119/mukherjee20a.html},
   year = {2020},
}

@article{Castelnovo2021,
   author = {Alessandro Castelnovo and Intesa Sanpaolo and Riccardo Crupi and Greta Greco and Daniele Regoli and Andrea Claudio Cosentini and Ilaria Giuseppina Penco},
   doi = {10.21203/RS.3.RS-1162350/V1},
   title = {The Zoo of Fairness Metrics in Machine Learning},
   url = {https://www.researchsquare.com https://www.researchsquare.com/article/rs-1162350/v1},
   year = {2021},
}

@article{Castelnovo2022,
   author = {Alessandro Castelnovo and Riccardo Crupi and Greta Greco and Daniele Regoli and Ilaria Giuseppina Penco and Andrea Claudio Cosentini},
   doi = {10.1038/s41598-022-07939-1},
   journal = {Scientific Reports 2022 12:1},
   title = {A clarification of the nuances in the fairness metrics landscape},
   url = {https://www.nature.com/articles/s41598-022-07939-1},
   year = {2022},
}

@article{Cunningham2020,
   author = {Padraig Cunningham and Sarah Jane Delany},
   doi = {10.1007/978-3-030-73959-1_2},
   journal = {Lecture Notes in Computer Science (including subseries Lecture Notes in Artificial Intelligence and Lecture Notes in Bioinformatics)},
   title = {Underestimation Bias and Underfitting in Machine Learning},
   url = {http://arxiv.org/abs/2005.09052 http://dx.doi.org/10.1007/978-3-030-73959-1_2},
   year = {2020},
}

@article{Ntoutsi2020,
   author = {Eirini Ntoutsi and Pavlos Fafalios and Ujwal Gadiraju and Vasileios Iosifidis and Wolfgang Nejdl and Maria Esther Vidal and Salvatore Ruggieri and Franco Turini and Symeon Papadopoulos and Emmanouil Krasanakis and Ioannis Kompatsiaris and Katharina Kinder-Kurlanda and Claudia Wagner and Fariba Karimi and Miriam Fernandez and Harith Alani and Bettina Berendt and Tina Kruegel and Christian Heinze and Klaus Broelemann and Gjergji Kasneci and Thanassis Tiropanis and Steffen Staab},
   doi = {10.1002/WIDM.1356},
   journal = {Wiley Interdisciplinary Reviews: Data Mining and Knowledge Discovery},
   title = {Bias in data-driven artificial intelligence systems—An introductory survey},
   url = {https://onlinelibrary.wiley.com/doi/full/10.1002/widm.1356},
   year = {2020},
}

@article{Fu2020,
   author = {Runshan Fu and Yan Huang and Param Vir Singh},
   doi = {10.1287/EDUC.2020.0215},
   journal = {INFORMS Tutorials in Operations Research},
   title = {Artificial Intelligence and Algorithmic Bias: Source, Detection, Mitigation, and Implications},
   url = {https://pubsonline.informs.org/doi/abs/10.1287/educ.2020.0215},
   year = {2020},
}

@article{Mehrabi2021,
   author = {Ninareh Mehrabi and Fred Morstatter and Nripsuta Saxena and Kristina Lerman and Aram Galstyan},
   doi = {10.1145/3457607},
   journal = {ACM Computing Surveys (CSUR)},
   title = {A Survey on Bias and Fairness in Machine Learning},
   url = {https://dl.acm.org/doi/10.1145/3457607},
   year = {2021},
}

@article{TrewinShari2019,
   author = {TrewinShari and BassonSara and MullerMichael and BranhamStacy and TreviranusJutta and GruenDaniel and HebertDaniel and LyckowskiNatalia and ManserErich},
   doi = {10.1145/3362077.3362086},
   journal = {AI Matters},
   title = {Considerations for AI fairness for people with disabilities},
   url = {https://dl.acm.org/doi/10.1145/3362077.3362086},
   year = {2019},
}

\end{document}